# Dark-Bright Soliton Bound States in a Microresonator

S. Zhang[1], T. Bi[1,2], G. N. Ghalanos[1,3], N. P. Moroney[1,3], L. Del Bino[1], and P. Del'Haye[1,2*]

[1]Max Planck Institute for the Science of Light, 91058 Erlangen, Germany.

[2]Department of Physics, Friedrich Alexander University Erlangen-Nuremberg, 91058 Erlangen, Germany.

[3]Blackett Laboratory, Imperial College London, SW7 2AZ, London, United Kingdom.

*Corresponding author: pascal.delhaye@mpl.mpg.de

**Abstract**

The recent discovery of dissipative Kerr solitons in microresonators has facilitated the development of fully coherent, chip-scale frequency combs. In addition, dark soliton pulses have been observed in microresonators in the normal dispersion regime. Here, we report bound states of mutually trapped dark-bright soliton pairs in a microresonator. The soliton pairs are generated seeding two modes with opposite dispersion but with similar group velocities. One laser operating in the anomalous dispersion regime generates a bright soliton microcomb, while the other laser in the normal dispersion regime creates a dark soliton via Kerr-induced cross-phase modulation with the bright soliton. Numerical simulations agree well with experimental results and reveal a novel mechanism to generate dark soliton pulses. The trapping of dark and bright solitons can lead to light states with the intriguing property of constant output power while spectrally resembling a frequency comb. These results can be of interest for telecommunication systems, frequency comb applications, ultrafast optics and soliton states in atomic physics.

**Introduction**

Microresonator-based frequency combs generated in monolithic high-Q microresonators have attracted significant research interest since initial demonstration in 2007 (*1*). In particular their small footprint and the possibility of chip-scale integration enables a large number of applications (*2–4*). Recently, the discovery of dissipative Kerr solitons in microresonators has been demonstrated as a source for low-noise and broadband frequency combs (*5*, *6*). Rich nonlinear dynamics in microresonators has been revealed in the past decade, including breather solitons (*7*, *8*), soliton crystals (*9*, *10*), Stokes solitons (*11*), Pockels solitons (*12*), laser cavity solitons (*13*), and dark solitons (*14*, *15*). In terms of applications, soliton microcombs have already been successfully used for optical frequency synthesizers (*16*), astronomy (*17*, *18*), optical coherent communications (*19*, *20*), laser-based light detection and ranging (*21–23*), and dual-comb spectroscopy (*24*, *25*), just to name a few.

Group velocity dispersion of microresonators plays a critical role in microcomb formation. Bright soliton generation in microresonators requires anomalous dispersion at the pump wavelengths while dark solitons can be observed when pumping in the normal dispersion regime. Recent research theoretically predicted the coexistence of bright and dark solitons in the regimes of normal (*26*), zero (*27*), and anomalous dispersion (*28*), when taking account of higher-order dispersion (third- and fourth-order). In addition, bichromatic pumping of microresonators at similar wavelengths has been studied for microcomb generation with the benefits of a thresholdless pump intensity and stabilization of the frequency comb repetition rate (*29–34*). More recently, bichromatic pumping has been demonstrated for the generation of dual orthogonally polarized microcombs (*35*, *36*) and spectral extension and synchronization of microcombs (*37*).

Dark-bright soliton pairs have been demonstrated in mode-locked lasers (*38*, *39*), however, there are no reports of their generation in microresonators. Here, we propose and experimentally demonstrate the generation of dark-bright soliton pairs in a single microresonator via bichromatic pumping, as illustrated in Fig.1. Two seed lasers at different wavelengths jointly pump a microresonator. One laser ($\lambda_1$, red) in the anomalous dispersion regime (referred to as primary pump) is used to generate a bright soliton microcomb. By exciting an optical mode in the normal dispersion regime, the second laser ($\lambda_2$, blue, referred to as auxiliary pump) passively forms dark soliton pulses through Kerr-induced cross-phase modulation (XPM) with the bright soliton. The dark soliton is trapped by the bright soliton in the time domain, resulting in a co-propagating pair of bright and dark solitons that are simultaneously emitted from the microresonator. A prerequisite for the generation of dark-bright soliton pairs is a similar free spectral range (FSR) in the spectral regions of both pump lasers. Numerical simulations agree excellently with experimental results and reveal a novel mechanism to generate dark soliton pulses. To the best of our knowledge, this is the first demonstration of simultaneous generation and trapping of bright and dark cavity solitons in passive resonators. The demonstrated technique could be useful for optical signal processing, future optical telecommunication (*40*) and atomic physics (*41*).

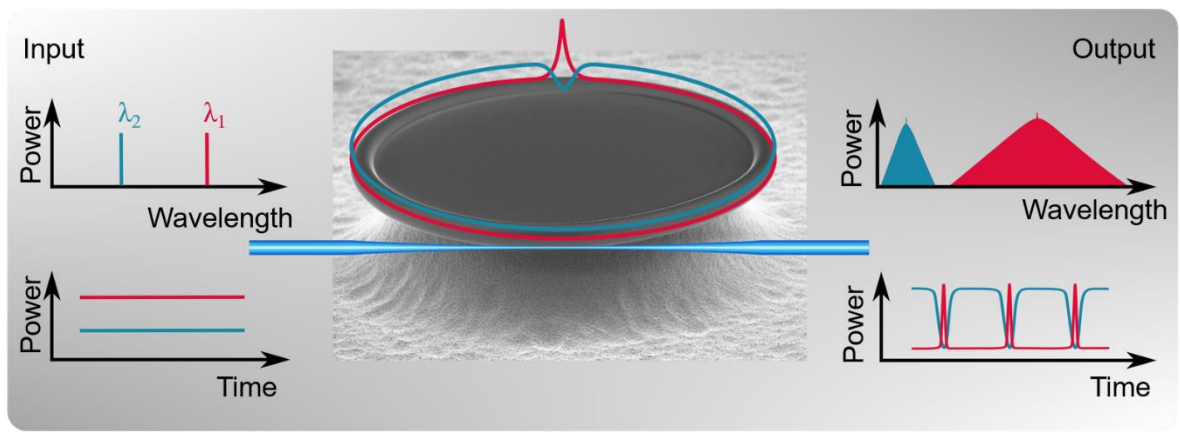

**Fig. 1. Scheme for dark-bright soliton pair generation in a microresonator with bichromatic pumping.** One seed laser ($\lambda_1$, red) operates in the anomalous dispersion regime, generating a bright soliton microcomb, while a second laser ($\lambda_2$, blue) excites a dark soliton in the normal dispersion regime. Both lasers operate in spectral regions with similar free spectral ranges. The dark soliton is synchronized with the bright soliton in the time domain, mediated by cross-phase modulation. Inset: Scanning electron microscope image of a silica microtoroid resonator used in the experiments.

**Dispersion of a microresonator**

Considering a passive whispering gallery mode resonator made of a dispersive medium with Kerr nonlinearity, the resonance frequencies of a mode family can be described as

$$\omega_{*,\mu} = \omega_{*,0} + D_{*,1}\mu + \frac{D_{*,2}}{2!}\mu^2 + \frac{D_{*,3}}{3!}\mu^3, \qquad (1)$$

where $\mu$ is the mode number offset from the pump mode at $\mu = 0$. $\omega_{*,\mu}$ is the resonance frequency of that mode, '*' refers to either the primary ('p') or auxiliary ('a') mode family, $D_{*,1}/2\pi$ is the FSR of the resonator at the primary ($D_{p,1}/2\pi$) or auxiliary ($D_{a,1}/2\pi$) pump mode ($\mu = 0$), and $D_{*,2}$, $D_{*,3}$ are coefficients of second- and third-order dispersion, respectively. Here, the two microresonator modes for the primary and auxiliary pumps can be either from the same mode family or different mode families (37). Figure 2A shows the normalized FSR mismatch $\gamma(\mu) = (D_{a,1}-D_{p,1}(\mu))/\Delta\omega_0$ (blue curve) as a function of the mode number $\mu$. The FSR mismatch $\gamma(\mu)$ is normalized to the full-width at half-maximum (FWHM) of the resonance $\Delta\omega_0$. Microresonators typically exhibit different dispersion over a broad spectral range. For example, silica microresonators exhibit normal dispersion (FSR decreases with optical frequency) at short wavelengths and anomalous dispersion (FSR increases with optical frequency) at long wavelengths. As a result, finding two optical modes of a microresonator at different wavelengths, and in opposite dispersion regimes with a similar FSR is actually feasible. The simulation (blue curve) plotted in Fig. 2A shows $\gamma(\mu)$ crossing zero when the relative mode number is around -8 ($\gamma \approx -0.02$ for $\mu = -8$). Thus, the group velocity of the primary pump mode at $\mu = -8$ is almost identical to that for the auxiliary pump mode.

**Numerical simulations of dark-bright soliton pairs**

Here, we perform numerical simulations based on two simultaneous, generalized Lugiato-Lefever equations (LLEs) (42–44), with additional XPM terms allowing interaction between primary and auxiliary fields, and considering the group velocity mismatch between the primary and auxiliary pump modes (35–37):

$$\frac{\partial \psi_p(\theta,\tau)}{\partial \tau} = -(1+i\alpha_p)\psi_p + i\left(|\psi_p|^2 + \sigma|\psi_a|^2\right)\psi_p - \sum_{n=2}^{N\geq 2}(-i)^{n+1}\frac{\beta_{p,n}}{n!}\frac{\partial^n \psi_p}{\partial \theta^n} + \gamma\frac{\partial \psi_p}{\partial \theta} + F_p, \qquad (2)$$

$$\frac{\partial \psi_a(\theta,\tau)}{\partial \tau} = -(1+i\alpha_a)\psi_a + i\left(|\psi_a|^2 + \sigma|\psi_p|^2\right)\psi_a - \sum_{n=2}^{N\geq 2}(-i)^{n+1}\frac{\beta_{a,n}}{n!}\frac{\partial^n \psi_a}{\partial \theta^n} - \gamma\frac{\partial \psi_a}{\partial \theta} + F_a, \qquad (3)$$

where $\tau$ is the slow time, normalized to twice the photon lifetime ($\tau_{ph}$) and $\theta$ is the azimuthal angle in a frame co-rotating at the average of the primary and auxiliary group velocities. $\psi_p(\theta,\tau)$, $\psi_a(\theta,\tau)$ are the intracavity primary and auxiliary field envelopes, respectively. $\alpha_p$ and $\alpha_a$ are the frequency detunings of the primary and auxiliary pump lasers with respect to their respective resonance frequencies and both normalized to half of $\Delta\omega_0$. $\sigma$ is the XPM coefficient (2/3 for orthogonal polarization and 2 for the same polarization assuming perfect spatial mode overlap, otherwise less). $\beta_{p,n}$, $\beta_{a,n}$ are the $n^{th}$-order dimensionless dispersion coefficients at the primary pump mode, normalized as $\beta_{p,n} = -2D_{p,n}/\Delta\omega_0$, $\beta_{a,n} = -2D_{a,n}/\Delta\omega_0$. $\gamma$ is the normalized group velocity mismatch between the primary and auxiliary pump modes. $F_p$, $F_a$ are the dimensionless external pump amplitudes. The LLE simulations are performed numerically using the split-step Fourier method. From Fig. 2A, the primary pump mode is selected at $\mu = -8$ to match the group velocity of the auxiliary mode. In the simulations, the auxiliary pump mode is fixed, and its dispersion parameters are $\beta_{a,2} = 0.2032$ and $\beta_{a,3} = 0.0022$. For the simulations in Fig. 2, the dispersion parameters of the primary pump mode are $\beta_{p,2} = -0.1190$, $\beta_{p,3} = 0.0023$, and $\gamma = -0.02$. The external pumping amplitudes for all of the simulations are $|F_a|^2 = 9$, $|F_p|^2 = 12$, and $\sigma = 1.5$.

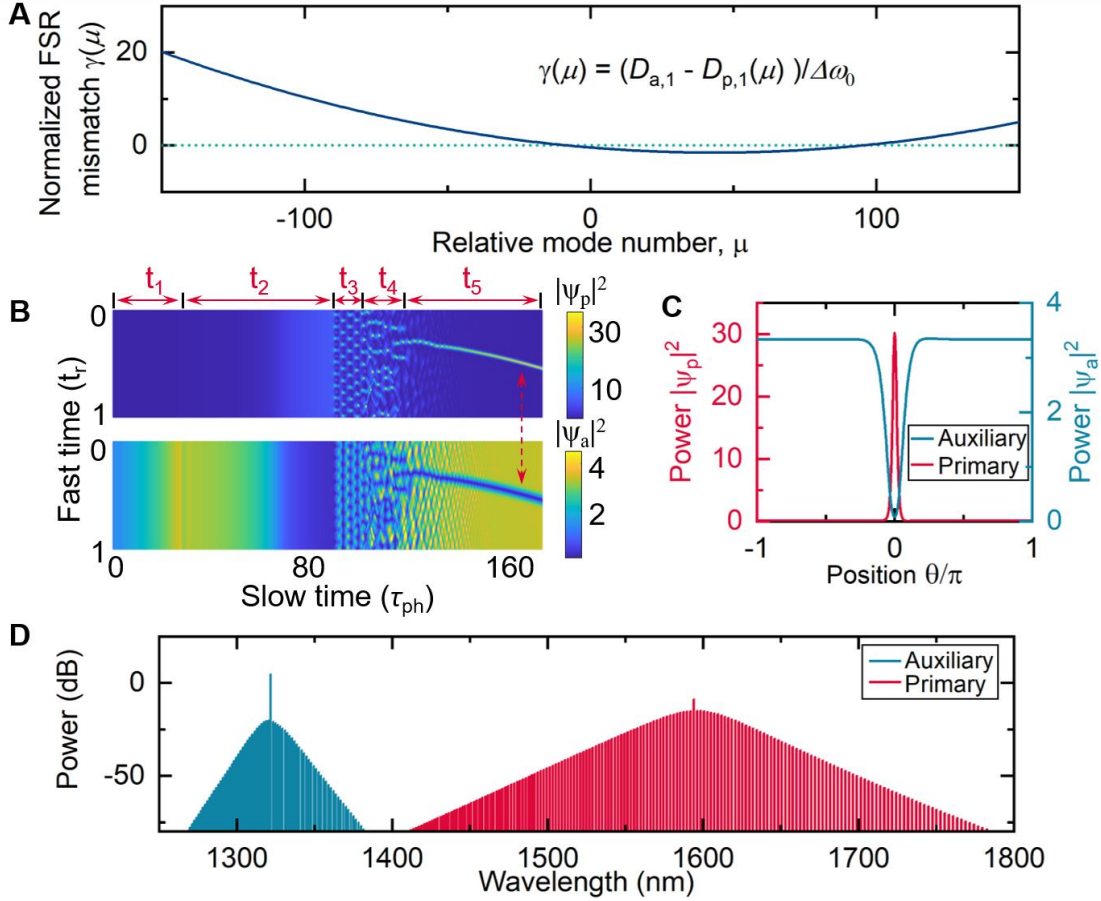

**Fig. 2. Numerical simulation of dark-bright soliton pair generation in a microresonator.** (**A**) Normalized group velocity mismatch $\gamma(\mu)$ (blue, solid line) between the auxiliary pump mode and the primary modes as a function of the mode number $\mu$. (**B**) Temporal waveform evolution of the primary bright soliton (upper panel) and auxiliary dark soliton (lower panel) when the primary laser seeds an optical mode at $\mu = -8$ (where the FSR matches the FSR around the auxiliary mode). $\tau_{ph}$: photon lifetime; $t_r$: roundtrip time. $|\psi_a|^2$ and $|\psi_p|^2$ are the corresponding intracavity powers. (**C**) Temporal waveforms of a bright soliton (red, left axis) and its induced dark soliton (blue, right axis) at the position marked with the red dashed arrow in (**B**). (**D**) Simulated intracavity optical spectrum corresponding to the temporal waveforms shown in (**C**).

Figure 2B depicts the generation and evolution of a dark-bright soliton pair. To match the simulation with experimental techniques, the auxiliary pump frequency is initially blue-detuned with respect to its resonance ($t_1$). Then the primary pump's frequency is scanned from the blue- to red-detuned side of its resonance until accessing stable soliton states ($t_2$ to $t_5$). The Kerr-induced resonance shift resulting from the rising intracavity primary power causes the decrease of the intracavity auxiliary power ($t_2$). When further tuning the primary laser into its resonance, the intracavity power will surpass the hyperparametric oscillation threshold power, and the primary field starts to generate Turing rolls ($t_3$) (*45*). Correspondingly, the auxiliary field also generates Turing rolls through Kerr XPM with the primary Turing rolls, however they are out of phase with each other (See Fig. S1 in the Supplementary Materials for details). When continuing to tune the primary laser into its resonance, the primary comb will enter a chaotic regime with low coherence ($t_4$), where both the intracavity primary and auxiliary waveform are unstable. Once the primary laser enters the red-detuned side of the resonance ($t_5$), a bright primary soliton is formed. Strikingly, a dark soliton is simultaneously generated for the auxiliary pump via the XPM interaction between the intracavity auxiliary field and the bright soliton. The dark soliton is stable and propagates at the same velocity as the bright soliton.

Figure 2C shows the simulated intracavity temporal waveforms of a single-soliton primary microcomb (red, left axis) and its induced auxiliary dark soliton pulse (blue, right axis). The soliton region is marked with a dashed arrow in Fig. 2B. The auxiliary dark soliton is trapped at the maximum of the bright pulse. The intracavity optical spectra corresponding to the temporal waveforms of Fig. 2C are shown in Fig. 2D. Compared with optical spectra of conventional microresonator dark solitons in the normal-dispersion regime (*14*, *15*), the spectral envelope shown in Fig. 2D does not exhibit "cat-ear" peaks neighboring the auxiliary pump. Thus, the temporal waveform of the generated dark soliton here has no low intensity oscillations in the dark soliton minima, in contrast to conventional dark solitons. Note that, when the primary microcomb is in a multi-soliton state, the resonator can also generate multiple dark-bright soliton pairs. See Fig. S2 in the Supplementary Materials for the simulation of multiple dark-bright soliton pairs under same parameters as in Fig. 2.

**Influence of the group velocity mismatch on the dark-bright soliton pair**

To investigate the influence of the group velocity mismatch on the dark soliton, we performed a number of simulations by changing the mode number that is pumped by the primary laser, while keeping the auxiliary laser seeding the same optical mode as in Fig.2. Figure 3 shows the simulation results for dark-bright soliton pairs at different group velocity mismatch when the primary soliton microcomb is in a single-soliton state. Figure 3A illustrates the simulated intracavity optical spectrum of the auxiliary dark soliton when the primary pump mode is at $\mu = -1$ with $\gamma = -0.41$, and the dispersion parameters of the primary pump mode are $\beta_{p,2} = -0.1028$, $\beta_{p,3} = 0.0023$. The corresponding temporal waveform of the dark soliton (blue) is shown in Fig. 3B with the red trace depicting the temporal waveform of the primary bright soliton. The spectra of the dark-bright soliton states can be found in Fig. S3 in the Supplementary Materials. Figures 3C and 3D show the simulation results when the primary pump mode is at $\mu = -11$ with $\gamma = 0.16$, and the dispersion parameters of the primary pump mode are $\beta_{p,2} = -0.1262$, $\beta_{p,3} = 0.0023$. Interestingly, the spectral and temporal profiles of the dark soliton change as seen in Figs. 3C and 3D. Compared with Fig. 2B, the spectra in Figs. 3A and 3C are more asymmetrical and irregular when the value of $\gamma$ is further from zero. The peak of the envelope of the auxiliary microcomb is shifted from the red (Fig. 3A) to the blue side (Fig. 3C) of the auxiliary pump frequency when the primary pump mode is changed from $\mu = -1$ to $-11$, which corresponds to a change of $\gamma$ from negative to positive. This asymmetry is also present in the time domain (as shown in Fig. 3B and D) where the relative amplitudes of the sidelobes-peaks of the dark soliton are changing when changing the sign of $\gamma$. Moreover, it is observed that the duration of the dark soliton increases when the absolute value of $\gamma$ increases. These results show that the group velocity mismatch $\gamma$ plays an important role in the temporal shape of the dark soliton.

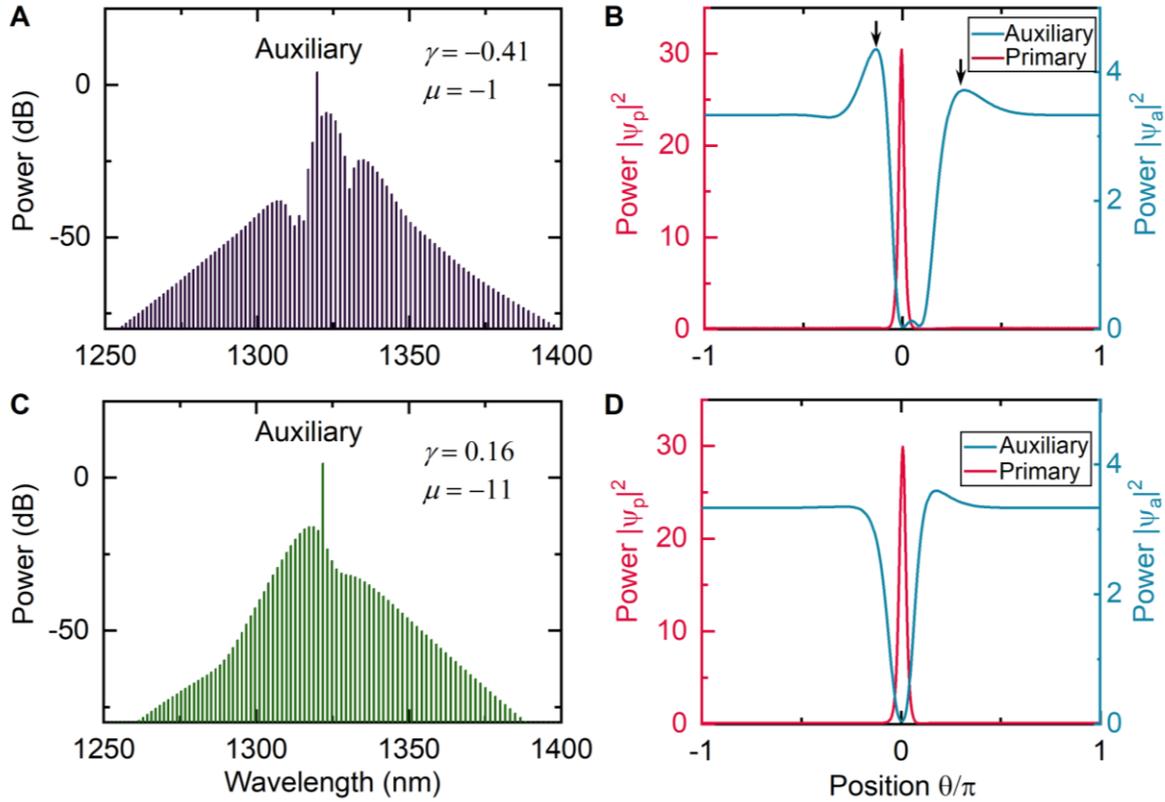

**Fig. 3. Dark-bright soliton pairs at different group velocity mismatch.** (**A**) Simulated optical spectrum and its corresponding temporal waveform (**B**) of the auxiliary dark soliton while the primary soliton microcomb is in a single-soliton state seeded at mode number $\mu$ = -1, with FSR mismatch $\gamma$ = - 0.41. (**C**) Simulated optical spectrum and its corresponding temporal waveform (**D**) of the auxiliary dark soliton when the primary soliton microcomb is seeded at $\mu$ = -11 with $\gamma$ = 0.16.

**Experimental demonstration of dark-bright soliton pairs**

The numerical predictions are verified with experiments using fused silica microtoroid resonators. Figure 4A shows the schematic of the experimental setup. Two external cavity diode lasers (ECDLs) are used for bichromatic pumping of a resonator. The primary laser at 1.5 μm generates a bright soliton microcomb, where the overall group velocity dispersion is anomalous; while the auxiliary laser at 1.3 μm passively stabilizes the circulating optical power inside the microresonator to assist bright soliton generation for the primary pump (*46*, *47*); meanwhile, generates a dark soliton pulse. A 235-μm-diameter fused silica microtoroid is used in the experiments with an FSR of 274 GHz. This particular microresonator is fabricated from a silicon wafer with a 6-μm layer of silicon dioxide (SiO$_2$) (*48*). These two lasers are combined with a wavelength division multiplexer (WDM) and evanescently coupled into the microresonator via a tapered optical fiber. Two fiber polarization controllers (PCs) are used to match the polarization of the two lasers. At the resonator output, part of the light is monitored by an optical spectrum analyzer (OSA). The other portion of the light is separated by another WDM to monitor each pump separately on two photodiodes (PD1 and PD2). An autocorrelator (AC) based on second-harmonic generation is used to measure the autocorrelation traces of the dark soliton pulse from the auxiliary pump. The primary mode family is carefully selected so that it not only supports the formation of a bright soliton, but also has a selection of modes with

FSRs that match the group velocity of the auxiliary mode. In the experiments, the 1.5-μm primary pump laser is widely tuned from 1560 nm to 1600 nm. Within this range, each mode from the primary mode family supports the formation of a bright soliton.

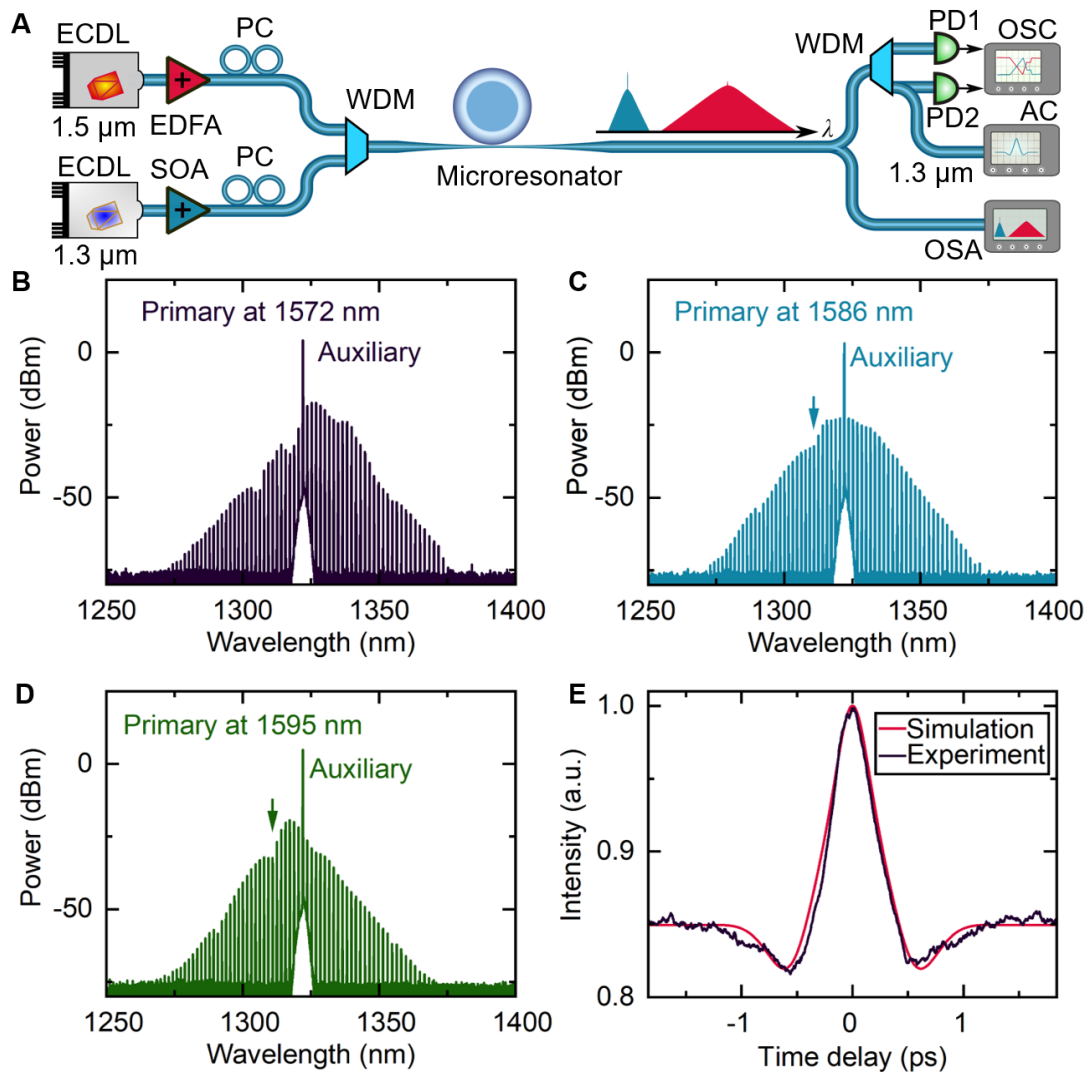

**Fig. 4. Experimental demonstration of a dark-bright soliton pairs.** (**A**) Setup used for the dark-bright soliton pair generation via bichromatic pumping. The 1.5-μm primary pump laser is used to generate a bright soliton, while the 1.3-μm auxiliary laser thermally stabilizes the resonator and at the same time generates a dark soliton via XPM with the bright soliton. ECDL: external cavity diode laser; EDFA: Erbium-doped fiber amplifier; SOA: Semiconductor optical amplifier; WDM: wavelength division multiplexer; PC: polarization controller; PD: photodetector; AC: autocorrelator; OSA: optical spectrum analyzer; OSC: oscilloscope. (**B**) (**C**) (**D**) Measured spectra of the dark soliton frequency combs for different central wavelength of the bright soliton. The bright soliton is seeded at 1572 nm in (**B**), 1586 nm in (**C**), and 1595 nm in (**D**). The arrows in (**C**) and (**D**) indicate a spectral dip induced by mode crossings. (**E**) Autocorrelation trace (black) of the dark soliton with the spectrum shown in (**B**) together with a calculated autocorrelation trace (red) of the dark soliton shown in the blue trace in Fig. 3**B**.

Figure 4B shows the dark soliton microcomb spectrum that is generated by the interplay of the primary bright soliton at 1572 nm seeded by 200 mW, together with 160 mW of auxiliary pump light at 1322 nm. See Fig. S4 in the Supplementary Materials for full spectrum including a spectrum of the single-soliton primary microcomb. We note that the dark soliton spectrum exhibits an irregular shape that agrees excellently with the simulated comb spectrum shown in Fig. 3A. In addition to the similar spectral envelope, the spectral peak of the auxiliary frequency comb also appears on the red side of the auxiliary pump, which is consistent with the simulation in Fig. 3A. To qualitatively explore the impact of the group velocity mismatch between the primary and auxiliary pump mode on the auxiliary microcomb in the experiments, the primary pump wavelength is changed to different modes between 1572 nm and 1595 nm while the auxiliary pump laser was held at a fixed mode at 1322 nm. As shown in Fig. 4C, the auxiliary microcomb assumes a symmetric shape around its pump when the primary microcomb is in a single-soliton state at 1586 nm. This is consistent with the simulated spectra shown in Fig 2D. In addition, the symmetric spectrum here suggests that the group velocity mismatch between the auxiliary mode and primary pump mode at 1586 nm is close to zero. The primary pump wavelength was further tuned to 1595 nm, which leads to the generation of an auxiliary frequency comb with a similar envelope as the one in Fig. 3C with a positive group velocity mismatch. The corresponding dark soliton spectrum is shown in Fig. 4D and exhibits a comb-envelope peak shifted to the blue side of the auxiliary pump frequency, in agreement with the simulation in Fig. 3C. The spectral defects (marked with arrows) that are present in both Figs 4C and 4D, are most likely generated by mode crossings with other mode families (*49*). See Fig. S5 in Supplementary Materials for measurements of spectra with multiple dark-bright soliton pairs.

The auxiliary combs presented in Fig. 4 agree well with the simulations in Fig. 2 and 3 in terms of their spectral envelopes and the dependence of the spectral peak on the group velocity mismatch. When the primary pump wavelength is tuned from short wavelength (1572 nm) to long wavelength (1595 nm), the group velocity mismatch changes from negative to positive, crossing zero at around 1586 nm. This is similar to the simulations in Fig. 3: when changing the sign of the group velocity mismatch from negative to positive, the spectral peak is shifted from the red to blue side of the auxiliary pump.

To further confirm our findings, we measure the autocorrelation traces of the generated auxiliary dark solitons. Figure 4E shows a normalized autocorrelation trace (black) of the dark soliton with a corresponding spectrum in Fig. 4B. For comparison, a simulated autocorrelation trace (red) of the dark soliton (blue trace shown in Fig. 3B) is overlaid. Both the experimental measurement and the simulation contain two dips on the sides of the autocorrelation peak, which are associated with the peaks shown in Fig. 3B (marked with arrows). Note that, the high background intensity of the autocorrelation trace is a result of the relatively short pulse duration of the dark soliton of ~ 500 fs.

**Discussion**

Conventional dark soliton formation in normal-dispersion microresonators requires the aid of mode interactions, which can be realized via mode coupling between different mode families in an overmoded microresonator (*14, 15*) or via mode interaction from an additional coupled microresonator (*50*). In contrast, the demonstrated method here generates dark solitons via the Kerr XPM with bright solitons, rather than the assistance of mode crossings. By pumping an optical mode with similar FSR to the primary mode, the local normal dispersion of the auxiliary mode is still preserved, and not modified via the synchronization of the bright and dark soliton, hence, enabling the generation of dark solitons (*37*). Compared with their generation in mode-locked lasers, the dark-bright soliton pairs demonstrated here can have higher repetition rates spanning from tens of GHz to THz depending on the diameter of the microresonator. In addition, the auxiliary pump mode can be selected from an orthogonally polarized mode family with respect to the primary pump mode, which enables the generation of orthogonally polarized dark-bright pulse pairs.

In summary, we proposed and experimentally demonstrated the generation of dark-bright soliton bound states in a single microresonator with two seed lasers: One laser operating in the anomalous dispersion regime generates a bright soliton, while the second laser seeds an optical mode in the normal dispersion regime with a similar group velocity to the first pump mode, to generate a dark soliton via XPM with the bright soliton. Operating the primary microcomb in a multi-soliton state, this technique can generate multiple dark-bright soliton pairs. The mutually trapped dark-bright soliton pairs enable waveforms with nearly constant output power in the time domain, which makes them less susceptible to perturbations and external noise in the optical systems. We believe the demonstrated technique could gain attention in ultrafast nonlinear optics and find potential applications in optical signal processing, future secure optical telecommunication (*40*), and atomic physics (*41*).

**References**

1. P. Del'Haye, A. Schliesser, O. Arcizet, T. Wilken, R. Holzwarth, T. J. Kippenberg, Optical frequency comb generation from a monolithic microresonator. *Nature* **450**, 1214–1217 (2007).
2. T. J. Kippenberg, R. Holzwarth, S. A. Diddams, Microresonator-Based Optical Frequency Combs. *Science* **332**, 555–559 (2011).
3. B. Stern, X. Ji, Y. Okawachi, A. L. Gaeta, M. Lipson, Battery-operated integrated frequency comb generator. *Nature* **562**, 401–405 (2018).
4. C. Xiang, J. Liu, J. Guo, L. Chang, R. N. Wang, W. Weng, J. Peters, W. Xie, Z. Zhang, J. Riemensberger, J. Selvidge, T. J. Kippenberg, J. E. Bowers, Laser soliton microcombs on silicon. *arXiv:2103.02725* (2021).
5. T. Herr, V. Brasch, J. D. Jost, C. Y. Wang, N. M. Kondratiev, M. L. Gorodetsky, T. J. Kippenberg, Temporal solitons in optical microresonators. *Nat. Photonics* **8**, 145–152 (2014).
6. T. J. Kippenberg, A. L. Gaeta, M. Lipson, M. L. Gorodetsky, Dissipative Kerr solitons in optical microresonators. *Science* **361**, eaan8083 (2018).
7. M. Yu, J. K. Jang, Y. Okawachi, A. G. Griffith, K. Luke, S. A. Miller, X. Ji, M. Lipson, A. L. Gaeta, Breather soliton dynamics in microresonators. *Nat. Commun.* **8**, 14569 (2017).
8. E. Lucas, M. Karpov, H. Guo, M. L. Gorodetsky, T. J. Kippenberg, Breathing dissipative solitons in optical microresonators. *Nat. Commun.* **8**, 736 (2017).
9. D. C. Cole, E. S. Lamb, P. Del'Haye, S. A. Diddams, S. B. Papp, Soliton crystals in Kerr resonators. *Nat. Photonics* **11**, 671–676 (2017).
10. M. Karpov, M. H. P. Pfeiffer, H. Guo, W. Weng, J. Liu, T. J. Kippenberg, Dynamics of soliton crystals in optical microresonators. *Nat. Phys.* **15**, 1071–1077 (2019).
11. Q.-F. Yang, X. Yi, K. Y. Yang, K. Vahala, Stokes solitons in optical microcavities. *Nat. Phys.* **13**, 53–57 (2017).
12. A. W. Bruch, X. Liu, Z. Gong, J. B. Surya, M. Li, C.-L. Zou, H. X. Tang, Pockels soliton microcomb. *Nat. Photonics* **15**, 21–27 (2021).

13. H. Bao, A. Cooper, M. Rowley, L. Di Lauro, J. S. Totero Gongora, S. T. Chu, B. E. Little, G.-L. Oppo, R. Morandotti, D. J. Moss, B. Wetzel, M. Peccianti, A. Pasquazi, Laser cavity-soliton microcombs. *Nat. Photonics* **13**, 384–389 (2019).
14. X. Xue, Y. Xuan, Y. Liu, P.-H. Wang, S. Chen, J. Wang, D. E. Leaird, M. Qi, A. M. Weiner, Mode-locked dark pulse Kerr combs in normal-dispersion microresonators. *Nat. Photonics* **9**, 594–600 (2015).
15. E. Nazemosadat, A. Fülöp, Ó. B. Helgason, P.-H. Wang, Y. Xuan, D. E. Leaird, M. Qi, E. Silvestre, A. M. Weiner, V. Torres-Company, Switching dynamics of dark-pulse Kerr frequency comb states in optical microresonators. *Phys. Rev. A* **103**, 013513 (2021).
16. D. T. Spencer, T. Drake, T. C. Briles, J. Stone, L. C. Sinclair, C. Fredrick, Q. Li, D. Westly, B. R. Ilic, A. Bluestone, N. Volet, T. Komljenovic, L. Chang, S. H. Lee, D. Y. Oh, M.-G. Suh, K. Y. Yang, M. H. P. Pfeiffer, T. J. Kippenberg, E. Norberg, L. Theogarajan, K. Vahala, N. R. Newbury, K. Srinivasan, J. E. Bowers, S. A. Diddams, S. B. Papp, An optical-frequency synthesizer using integrated photonics. *Nature* **557**, 81–85 (2018).
17. E. Obrzud, M. Rainer, A. Harutyunyan, M. H. Anderson, J. Liu, M. Geiselmann, B. Chazelas, S. Kundermann, S. Lecomte, M. Cecconi, A. Ghedina, E. Molinari, F. Pepe, F. Wildi, F. Bouchy, T. J. Kippenberg, T. Herr, A microphotonic astrocomb. *Nat. Photonics* **13**, 31–35 (2019).
18. M.-G. Suh, X. Yi, Y.-H. Lai, S. Leifer, I. S. Grudinin, G. Vasisht, E. C. Martin, M. P. Fitzgerald, G. Doppmann, J. Wang, D. Mawet, S. B. Papp, S. A. Diddams, C. Beichman, K. Vahala, Searching for exoplanets using a microresonator astrocomb. *Nat. Photonics* **13**, 25–30 (2019).
19. P. Marin-Palomo, J. N. Kemal, M. Karpov, A. Kordts, J. Pfeifle, M. H. P. Pfeiffer, P. Trocha, S. Wolf, V. Brasch, M. H. Anderson, R. Rosenberger, K. Vijayan, W. Freude, T. J. Kippenberg, C. Koos, Microresonator-based solitons for massively parallel coherent optical communications. *Nature* **546**, 274–279 (2017).
20. A. Fülöp, M. Mazur, A. Lorences-Riesgo, Ó. B. Helgason, P.-H. Wang, Y. Xuan, D. E. Leaird, M. Qi, P. A. Andrekson, A. M. Weiner, V. Torres-Company, High-order coherent communications using mode-locked dark-pulse Kerr combs from microresonators. *Nat. Commun.* **9**, 1598 (2018).
21. M.-G. Suh, K. J. Vahala, Soliton microcomb range measurement. *Science* **359**, 884–887 (2018).
22. P. Trocha, M. Karpov, D. Ganin, M. H. P. Pfeiffer, A. Kordts, S. Wolf, J. Krockenberger, P. Marin-Palomo, C. Weimann, S. Randel, W. Freude, T. J. Kippenberg, C. Koos, Ultrafast optical ranging using microresonator soliton frequency combs. *Science* **359**, 887–891 (2018).
23. N. Kuse, M. E. Fermann, Frequency-modulated comb LIDAR. *APL Photonics* **4**, 106105 (2019).
24. M.-G. Suh, Q.-F. Yang, K. Y. Yang, X. Yi, K. J. Vahala, Microresonator soliton dual-comb spectroscopy. *Science* **354**, 600–603 (2016).
25. A. Dutt, C. Joshi, X. Ji, J. Cardenas, Y. Okawachi, K. Luke, A. L. Gaeta, M. Lipson, On-chip dual-comb source for spectroscopy. *Sci. Adv.* **4**, e1701858 (2018).
26. P. Parra-Rivas, D. Gomila, L. Gelens, Coexistence of stable dark- and bright-soliton Kerr combs in normal-dispersion resonators. *Phys. Rev. A* **95**, 053863 (2017).
27. J. H. Talla Mbé, C. Milián, Y. K. Chembo, Existence and switching behavior of bright and dark Kerr solitons in whispering-gallery mode resonators with zero group-velocity dispersion. *Eur. Phys. J. D.* **71**, 196 (2017).

28. J. H. T. Mbé, Y. K. Chembo, Coexistence of bright and dark cavity solitons in microresonators with zero, normal, and anomalous group-velocity dispersion: a switching wave approach. *J. Opt. Soc. Am. B* **37**, A69–A74 (2020).
29. D. V. Strekalov, N. Yu, Generation of optical combs in a whispering gallery mode resonator from a bichromatic pump. *Phys. Rev. A* **79**, 041805 (2009).
30. T. Hansson, S. Wabnitz, Bichromatically pumped microresonator frequency combs. *Phys. Rev. A* **90**, 013811 (2014).
31. H. Taheri, A. B. Matsko, L. Maleki, Optical lattice trap for Kerr solitons. *Eur. Phys. J. D.* **71**, 153 (2017).
32. C. Bao, P. Liao, A. Kordts, L. Zhang, M. Karpov, M. H. P. Pfeiffer, Y. Cao, Y. Yan, A. Almaiman, G. Xie, A. Mohajerin-Ariaei, L. Li, M. Ziyadi, S. R. Wilkinson, M. Tur, T. J. Kippenberg, A. E. Willner, Dual-pump generation of high-coherence primary Kerr combs with multiple sub-lines. *Opt. Lett.* **42**, 595–598 (2017).
33. W. Wang, S. T. Chu, B. E. Little, A. Pasquazi, Y. Wang, L. Wang, W. Zhang, L. Wang, X. Hu, G. Wang, H. Hu, Y. Su, F. Li, Y. Liu, W. Zhao, Dual-pump Kerr Micro-cavity Optical Frequency Comb with varying FSR spacing. *Sci. Rep.* **6**, 28501 (2016).
34. R. J. Weiblen, I. Vurgaftman, Bichromatic pumping in mid-infrared microresonator frequency combs with higher-order dispersion. *Opt. Express* **27**, 4238–4260 (2019).
35. C. Bao, P. Liao, A. Kordts, L. Zhang, A. Matsko, M. Karpov, M. H. P. Pfeiffer, G. Xie, Y. Cao, A. Almaiman, M. Tur, T. J. Kippenberg, A. E. Willner, Orthogonally polarized frequency comb generation from a Kerr comb via cross-phase modulation. *Opt. Lett.* **44**, 1472–1475 (2019).
36. R. Suzuki, S. Fujii, A. Hori, T. Tanabe, Theoretical Study on Dual-Comb Generation and Soliton Trapping in a Single Microresonator with Orthogonally Polarized Dual Pumping. *IEEE Photonics J.* **11**, 6100511 (2019).
37. S. Zhang, J. M. Silver, T. Bi, P. Del'Haye, Spectral extension and synchronization of microcombs in a single microresonator. *Nat. Commun.* **11**, 6384 (2020).
38. Q. Ning, S. Wang, A. Luo, Z. Lin, Z. Luo, W. Xu, Bright–Dark Pulse Pair in a Figure-Eight Dispersion-Managed Passively Mode-Locked Fiber Laser. *IEEE Photonics J.* **4**, 1647–1652 (2012).
39. G. Shao, Y. Song, L. Zhao, D. Shen, D. Tang, Soliton-dark pulse pair formation in birefringent cavity fiber lasers through cross phase coupling. *Opt. Express* **23**, 26252–26258 (2015).
40. Y. S. Kivshar, B. Luther-Davies, Dark optical solitons: physics and applications. *Phys. Rep.* **298**, 81–197 (1998).
41. C. Becker, S. Stellmer, P. Soltan-Panahi, S. Dörscher, M. Baumert, E.-M. Richter, J. Kronjäger, K. Bongs, K. Sengstock, Oscillations and interactions of dark and dark–bright solitons in Bose–Einstein condensates. *Nat. Phys.* **4**, 496–501 (2008).
42. L. A. Lugiato, R. Lefever, Spatial Dissipative Structures in Passive Optical Systems. *Phys. Rev. Lett.* **58**, 2209–2211 (1987).
43. S. Coen, H. G. Randle, T. Sylvestre, M. Erkintalo, Modeling of octave-spanning Kerr frequency combs using a generalized mean-field Lugiato-Lefever model. *Opt. Lett.* **38**, 37–39 (2013).
44. Y. K. Chembo, C. R. Menyuk, Spatiotemporal Lugiato-Lefever formalism for Kerr-comb generation in whispering-gallery-mode resonators. *Phys. Rev. A* **87**, 053852 (2013).
45. A. Coillet, I. Balakireva, R. Henriet, K. Saleh, L. Larger, J. M. Dudley, C. R. Menyuk, Y. K. Chembo, Azimuthal Turing Patterns, Bright and Dark Cavity Solitons in Kerr Combs Generated With Whispering-Gallery-Mode Resonators. *IEEE Photonics J.* **5**, 6100409 (2013).

46. S. Zhang, J. M. Silver, L. D. Bino, F. Copie, M. T. M. Woodley, G. N. Ghalanos, A. Ø. Svela, N. Moroney, P. Del'Haye, Sub-milliwatt-level microresonator solitons with extended access range using an auxiliary laser. *Optica* **6**, 206–212 (2019).
47. H. Zhou, Y. Geng, W. Cui, S.-W. Huang, Q. Zhou, K. Qiu, C. Wei Wong, Soliton bursts and deterministic dissipative Kerr soliton generation in auxiliary-assisted microcavities. *Light Sci. Appl.* **8**, 50 (2019).
48. S. Zhang, J. M. Silver, J. M. Silver, X. Shang, L. D. Bino, L. D. Bino, N. M. Ridler, P. Del'Haye, Terahertz wave generation using a soliton microcomb. Opt. Express 27, 35257–35266 (2019).
49. T. Herr, V. Brasch, J. D. Jost, I. Mirgorodskiy, G. Lihachev, M. L. Gorodetsky, T. J. Kippenberg, Mode Spectrum and Temporal Soliton Formation in Optical Microresonators. *Phys. Rev. Lett.* **113**, 123901 (2014).
50. X. Xue, Y. Xuan, P.-H. Wang, Y. Liu, D. E. Leaird, M. Qi, A. M. Weiner, Normal-dispersion microcombs enabled by controllable mode interactions. *Laser Photonics Rev*. **9**, L23–L28 (2015).

**Acknowledgments**

**Funding:** This work was supported by the European Union's H2020 ERC Starting Grant "CounterLight" 756966, the H2020 Marie Sklodowska-Curie COFUND "Multiply" 713694, the Marie Curie Innovative Training Network "Microcombs" 812818, the Max Planck Society, and the Engineering and Physical Sciences Research Council (EPSRC) via the CDTs for Applied Photonics, Controlled Quantum Dynamics, and Quantum Systems Engineering.

**Author contributions:** S.Z. designed the experiments, performed the measurements and numerical simulations. S.Z., T. B., and P.D.H. analyzed the data. All co-authors contributed to the manuscript.

**Competing interests:** The authors declare no competing interests.

**Data and materials availability:** All data needed to evaluate the conclusions in the paper are present in the paper and/or the Supplementary Materials. Additional data related to this paper may be requested from the authors.

# Supplementary Materials for Dark-Bright Soliton Bound States in a Microresonator

Shuangyou Zhang, Toby Bi, George N. Ghalanos, Niall P. Moroney, Leonardo Del Bino, and Pascal Del'Haye

*Corresponding author. Email: pascal.delhaye@mpl.mpg.de

**Numerical simulation of dark-bright Turing rolls**

Figure S1A shows the numerical simulations of intracavity waveforms of the primary Turing rolls (red trace) and the induced auxiliary Turing rolls (blue), while the primary pump laser operates in a Turing pattern regime. Interestingly, they are out of phase with each other. Figure S1B shows the corresponding optical spectra.

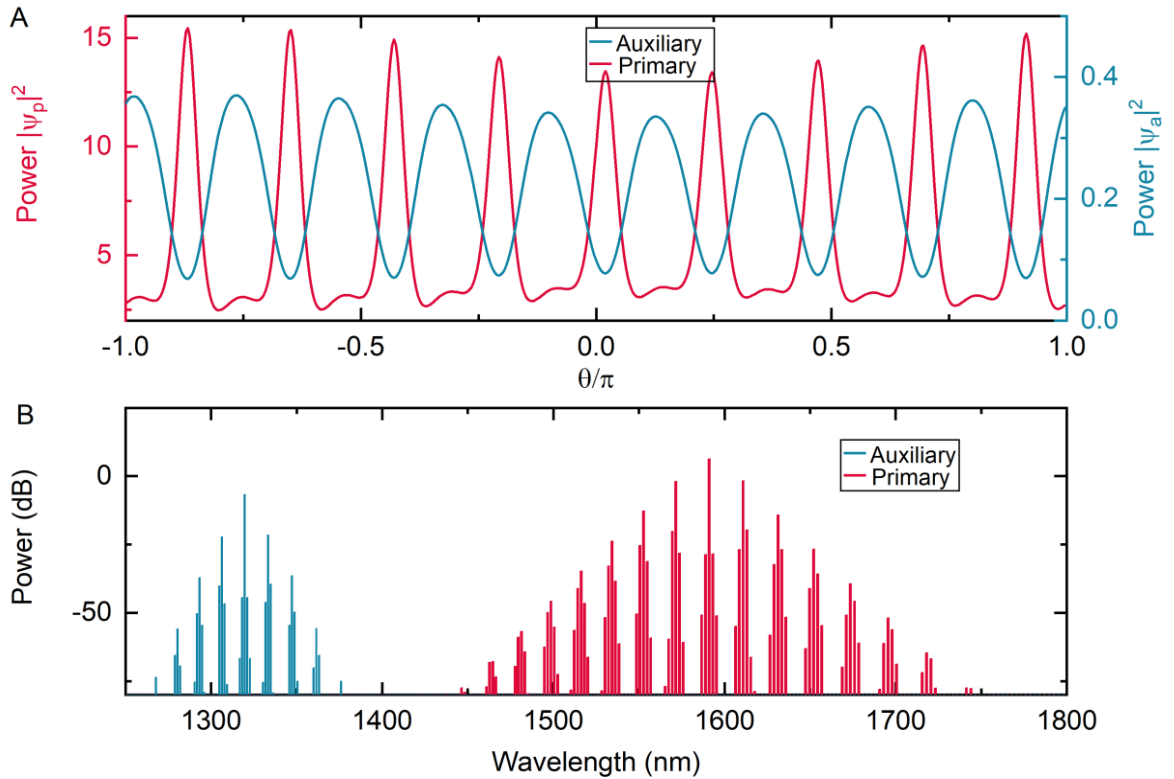

**Fig. S1.** Numerical simulations of the primary microcomb in a Turing pattern regime. (**A**) Intracavity waveforms of the primary Turing rolls (red trace) and the induced auxiliary Turing rolls (blue). $|\psi_a|^2$ and $|\psi_p|^2$ are the corresponding intracavity powers as a function of the position in the cavity given by $\theta/\pi$. (**B**) Corresponding optical spectra of the waveforms.

**Numerical simulation of multiple dark-bright soliton pairs in a microresonator**

Figure S2 shows numerical simulations of the generation of multiple dark-bright soliton pairs when the primary microcomb is in a multi-soliton ("soliton crystal") state. The simulation parameters are same as those in Fig. 2 of the main text. Figure S2A shows the intracavity optical spectrum of multiple dark-bright soliton pairs. The corresponding temporal waveforms are shown in Figure S2B. Figure S2C shows the temporal evolution of the waveforms of the two dark-bright soliton pairs. The numerical simulation shows the existence of two dark solitons that are simultaneously generated and trapped by the bright solitons.

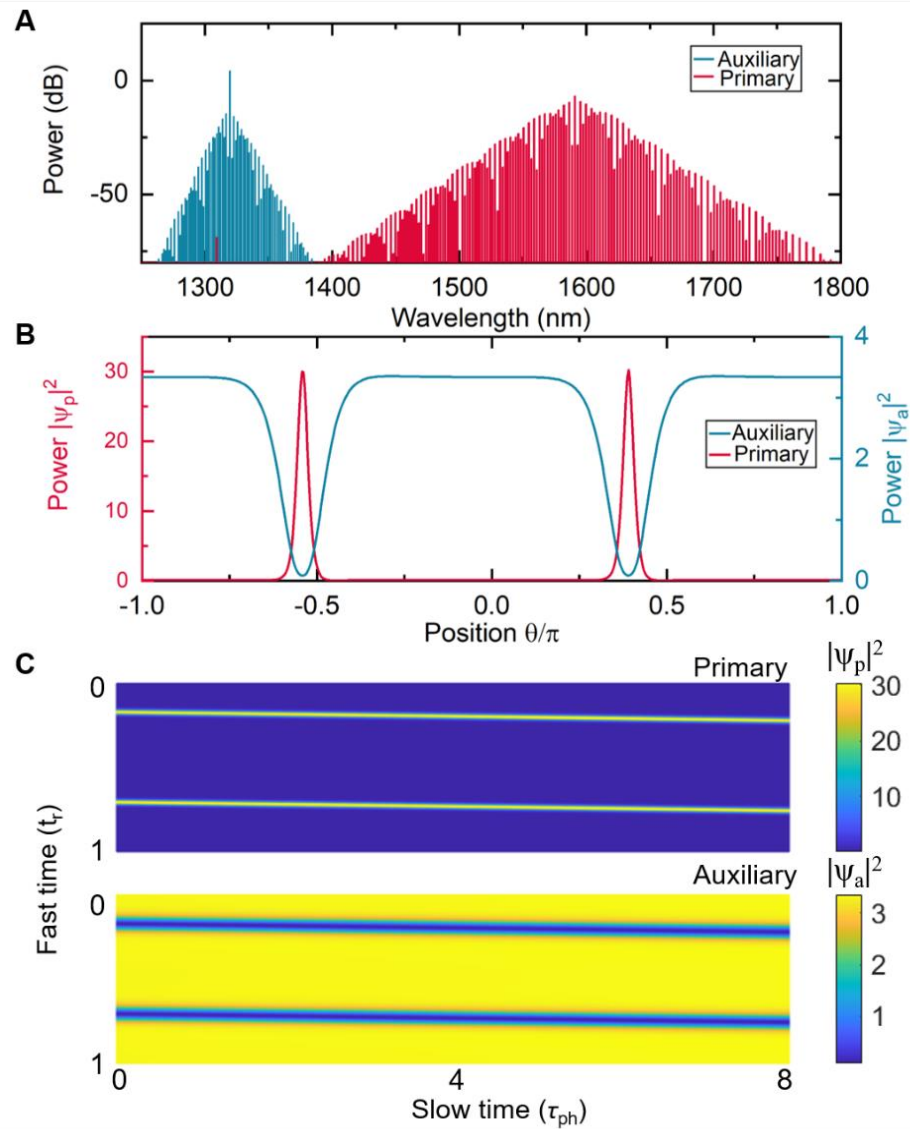

**Fig. S2.** (**A**) Simulated intracavity optical spectrum with the primary microcomb in a two-soliton state. (**B**) Temporal waveforms of two bright solitons (red, left axis) and their induced dark solitons (blue, right axis), corresponding to the optical spectrum shown in (**A**). (**C**) Temporal waveform evolution of the primary bright solitons (upper panel) and auxiliary dark solitons (lower panel).

**Cross-phase modulation (XPM) interaction between the dark and bright soliton**

As discussed and shown in Fig. 2 and 3 of the main text, the influence of the primary bright soliton on the auxiliary field results in a dark soliton pulse via XPM. In the same way, the generated dark soliton also has an influence on the primary bright soliton. Figure S3A shows the temporal intracavity waveforms of the bright and dark soliton (same plot as in Fig. 3B in the main text), and Figure S3B shows the zoomed-in part of the bright soliton pulse pedestal. There are two peaks on each side of the bright soliton, induced by the corresponding two peaks of the dark soliton at the same position within the resonator. As shown in Fig. S3C, this interaction from the dark soliton causes a slight change in the spectral envelope of the bright soliton, which leads to a slight divergence from a $sech^2$ shape around the primary pump. In addition, a spatiotemporal oscillation is induced by the beat signal between the primary pump and the dispersive wave around 1335 nm.

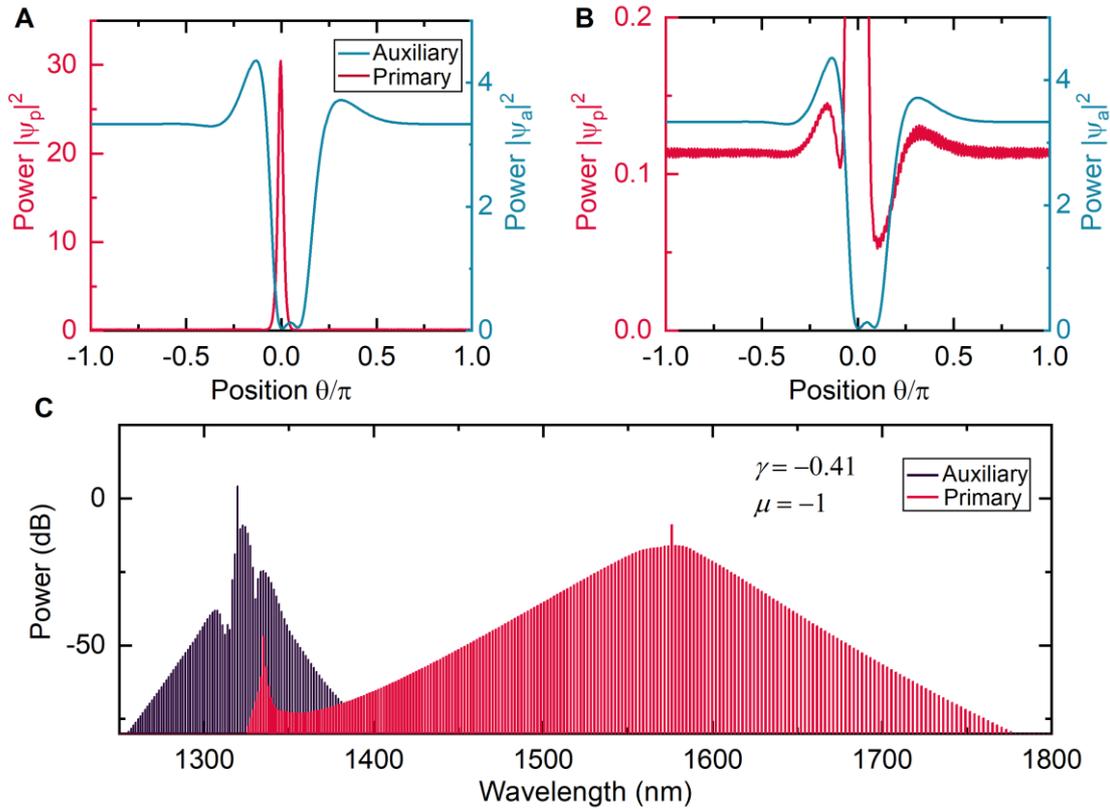

**Fig. S3.** (**A**) Simulated intracavity waveforms of a dark-bright soliton pair (same plot shown in Fig. 3A of the main text). (**B**) Enlarged view of the bright soliton pedestal. (**C**) Simulated optical spectrum of the dark-bright soliton pair, corresponding to the temporal waveforms in (**A**).

**Experimental spectra of the dark-bright soliton pair at different primary pump wavelengths**

Figure S4 shows the experimental spectra of a dark-bright soliton pair when the primary microcomb is in a single-soliton state and the primary pump laser is tuned to different wavelength: 1572 nm (upper panel), 1586 nm (middle panel), and 1595 nm (lower panel). The results plotted here correspond to the plots in Figs. 4B~D in the main text, including the spectra of the primary microcomb.

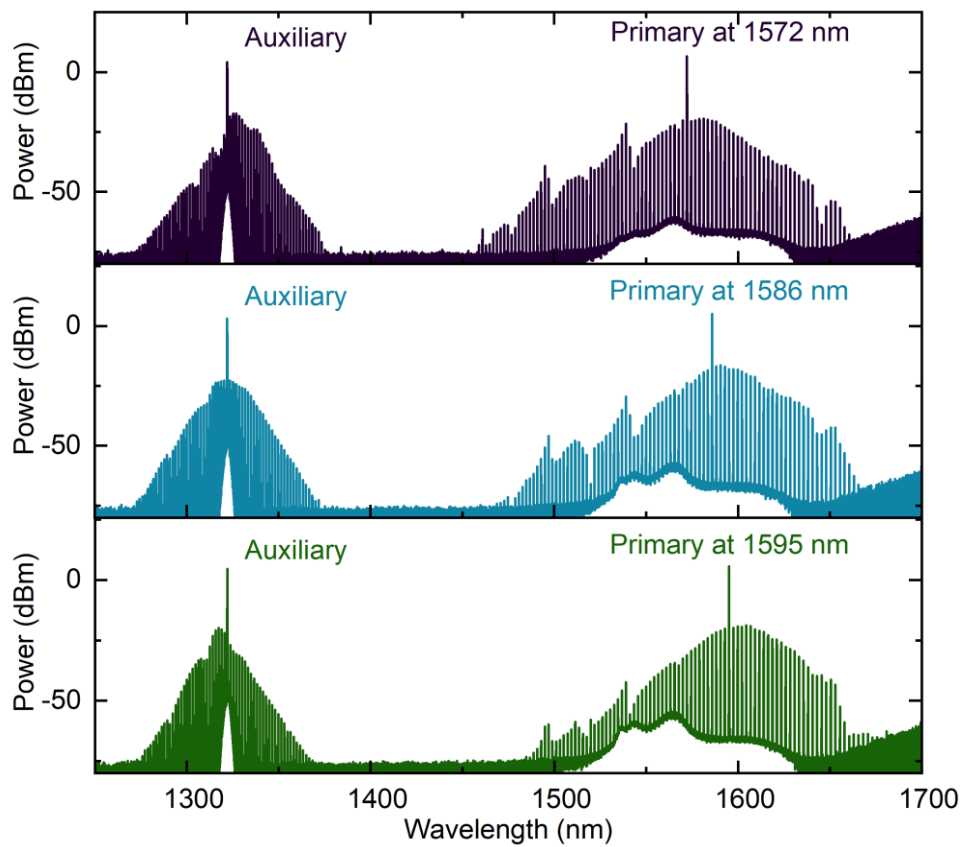

**Fig. S4.** Measured spectra of the auxiliary and primary frequency combs when the primary microcomb is in a single-soliton state and the primary pump laser operates at 1572 nm (upper panel), 1586 nm (middle panel), and 1595 nm (lower panel). The results here correspond to the results in Fig. 4 in the main text and include the spectra of the primary microcombs.

**Experimental demonstration of multiple dark-bright soliton pairs**

By detuning the primary laser frequency within its resonance, the primary microcomb can be deterministically changed between different soliton regimes that correspond to different numbers of solitons circulating within the resonator. When the primary comb is in a multi-soliton state, multiple bright solitons can be generated by the microresonator. As shown in Fig. S2B, the multiple bright solitons will induce and trap multiple dark solitons, resulting in multiple dark-bright soliton pairs. Figure S5 shows the measured spectra of multiple dark-bright soliton pairs with the primary pump operating at a wavelength of 1572 nm, under the same experimental condition as Fig. 4B in the main text.

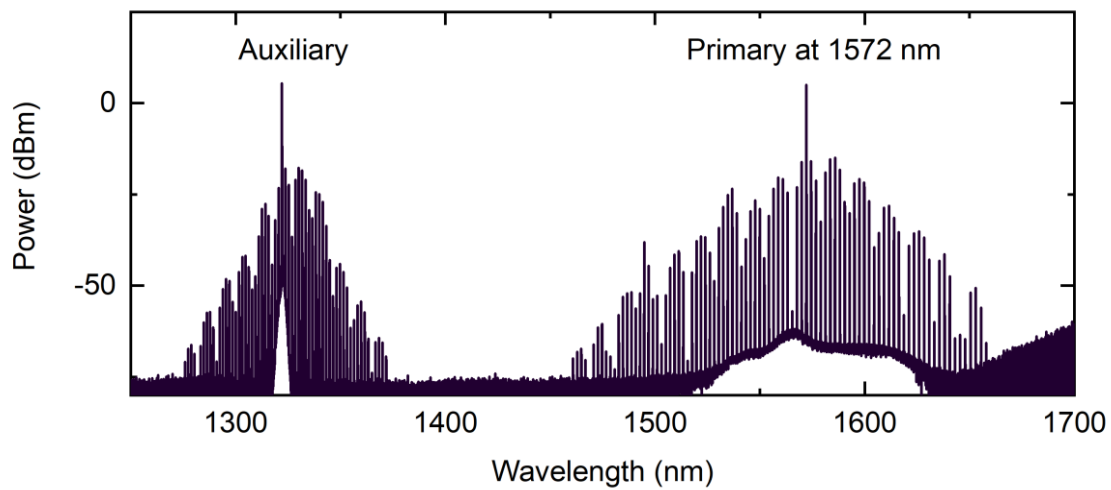

**Fig. S5.** Measured spectra of multiple dark-bright soliton pairs with the primary microcomb in a multi-soliton state and the primary pump laser operating at a wavelength of 1572 nm.